\documentclass{article}
\usepackage[latin9]{inputenc}
\usepackage{geometry}
\usepackage{amsmath}
\usepackage{amssymb}
\usepackage[unicode=true,pdfusetitle,
 bookmarks=true,bookmarksnumbered=false,bookmarksopen=false,
 breaklinks=false,pdfborder={0 0 1},backref=section,colorlinks=false]
 {hyperref}
\usepackage{breakurl}

\makeatletter

\usepackage{float}
\usepackage{graphicx}
\usepackage{grffile}
\usepackage{breakurl}

\providecommand{\tabularnewline}{\\}

\usepackage{fullpage}\usepackage[justification=centering]{caption}\usepackage[all]{hypcap}\usepackage{microtype}

\makeatother

\begin{document}

\title{Wave functions of the Q.Q interaction in terms of unitary $9j$ coefficients}

\author{Larry Zamick and Matthew Harper\\
 \textit{Department of Physics and Astronomy, Rutgers University,
Piscataway, New Jersey 08854}\\
 }

\date{\today}

\maketitle
 
\begin{abstract}
We obtain wave functions for two protons and two neutrons in the $g_{9/2}$
shell expressed as column vectors with amplitudes $D(J_{p},J_{n})$.
When we use a quadrupole-quadrupole interaction (Q.Q) we get, in many
cases, a very strong overlap with wave functions given by a single
set of unitary $9j$ coefficients -- U$9j=\langle(jj)^{2j}(jj^{J_{B}}\vert(jj)^{J_{p}}(jj)^{J_{n}})^{I}\rangle$.
Here $J_{B}=9$ for even $I$ $T$=0 states. For both even and odd
T=1 states we take J$_{B}$ equal to 8 whilst for odd I ,T=0 we take
JB to be 7. We compare the Q.Q results with those of a more realistic
interaction. 
\end{abstract}

\section{Introduction}

\indent \indent In previous works, the problem of maximum-$J$ $T=0$
pairing was addressed\cite{ZamickEscuderos,Kleszyk,HertzZamickKleszyk}
and comparisons were made with $J=0$ $T=1$ pairing. In the course
of these works it was found that to an excellent approximation some
wave functions of the maximum-$J$ pairing Hamiltonian were very close
to single sets of unitary 9-$j$ coefficients. In this work we wish
to disengage this simple result from the complexities of the maximum
$J$ pairing problem. To this end, we consider better Hamiltonians
than max-$J$ and show that the results hold there as well, and we
will make comparisons with wave functions obtained from the simple
Q.Q interaction as well as a more realistic interaction CCGI \cite{Coraggio}.
The calculations are for two protons and two neutrons in the $g_{9/2}$
shell i.e. $^{96}$Cd.

\section{Overlaps of Q.Q and CCGI with U$9j$}

\indent \indent To make the comparisons of the two interactions
easier we add constants so that the $J$=0 matrix elements are zero
for both interactions. The ten matrix elements from $J=0^{+}$ to
$J=9^{+}$ are then:

Q.Q : 0, 0.1222, 0.3485, 0.6515, 0.9848, 1.2879, 1.4849, 1.4849, 1.1818,
0.4546

CCGI: 0, 0.8290, 1.6500, 1.8770, 2.2170, 2.3018, 1.6049, 2.3830, 1.8019,
2.5270, 0.9150\\
 The $J=0$ values are -1.0000 and -2.3170 MeV respectively. Of course
the Q.Q interaction can be multiplied by a positive constant without
changing the overlaps. We first compare the overlaps of wave functions
obtained with the popular Q.Q interactions with wave functions that
are basically single sets of U$9j$ coefficients. This interaction
has the nice feature of having attractive $J=0^{+}$, 1$^{+}$ and
9$^{+}$ two-body matrix elements. For even $I$ we compare the yrast
state wave functions of Q.Q with those of the U$9j$'s, i.e. $N\langle(jj)^{J_{max}}(jj)^{J_{B}}|(jj)^{J_{p}}(jj)^{J_{n}}\rangle^{I}$,
where in the $g_{9/2}$ shell $J_{max}=2j=9$. To compare with yrast
$T=0$ even $I$ states of Q.Q we take $J_{B}=2j=9$. The normalization
$N$ is close to $\sqrt{2}$. More precisely, $N(9)^{-2}=1/2-1/2\langle(jj)^{9}(jj)^{9}|(jj)^{9}(jj)^{9}\rangle$.\cite{ZamickEscuderos}
\\
 We show in Table 1 the following overlaps($\psi$, U$9j$) for both
Q.Q and CCGI\cite{Coraggio}. \\
 Note the very strong overlaps for $I=0^{+},2^{+},4^{+}$ and $6^{+}$
and then the sudden drop to almost zero overlap for $I=8^{+}$ and
the small overlap of 0.3635 for $I=10^{+}$. In selected cases we
consider overlaps with the next excited states e.g. $T=0$ $I=8^{+}$
and $10^{+}$. The results were 0.9505 for $I=8^{+}$ and 0.8540 for
$I=10^{+}$. In other words for $I=8^{+}$ we can, to a good approximation
identify the simple U$9j$ wave function with the first exited state
rather than the ground state. For $I=10^{+}$ there is fragmentation.
For $I=12^{+}$ and $14^{+}$ we get poor overlaps of 0.6586 and 0.8374
respectively. However, for $I=16^{+}$ we get a perfect overlap. But
this case is trivial. There is only one $I=16^{+}$state with $J_{p}=8$
and $J_{n}=8$. \\
 We next briefly consider the other $(J,T)$ combinations. If the
wave function amplitudes are $D(J_{p},J_{n})^{I}$ then we have $D(J_{n},J_{p})=(-1)^{(I+T)}$
$D(J_{p},J_{n})$ where $T$ is the isospin. Thus, for even $I$ $T=1$
and for odd $I$ $T=1$ we take $J_{B}=8$ whilst for odd $I$ $T=0$
we take $J_{B}=7$. Note that there are no $I=0$ $T=1$ states in
this model space and that all $I=1$ states have isospin $T=1$. \\
 Referring to Table I we here note the $(J,T)$ states of these other
combinations with overlap\\
 greater than 0.9: (2,1), (4,1), (6,1), (3,0), (5,0), (3,1), (5,1)\\
 There is also the special case (14,1), here there is a perfect overlap
because there is only one state of this configuration. Except for
the last case, we get best overlaps for the lowest angular momenta.
Note that we cannot associate U$9j$'s with $T=2$ states. Those are
double analogs of states of four identical nucleons and the wave functions
are uniquely constrained by the Pauli Principle. \\
 By choosing $J_{B}=8$, we get a spectacular overlap of 0.9990 for
the lowest $I=1^{+}$ $T=1$ state. Equally impressive for the $I=3^{+}$
$T=1$ state, the overlap is 0.9997. Choosing $J_{B}=7$ we get for
the $I=3^{+}$ $T=0$ an overlap of 0.9939 and for $I=5^{+}$ $T=0$
we get 0.9125. For values of $I$ beyond $I=5$ however, things begin
to erode just as they do for large even $I$. \\
 In general, we get better overlaps with Q.Q than we do with the more
realistic CCGI. The values for $T=0$ $I=0$ are respectively 0.9996
and 0.9451. For $I=1$ $T=1$ they are 0.9990 and 0.8369. This may
be due in part to the fact that Q.Q has a more attractive $J_{max}$
matrix element than CCGI does. \\
 But there are some surprises. For $J=12^{+}$ and $14^{+}$ we get
much better results for CCGI than for Q.Q. The (Q.Q,CCGI) values are
(0.6590,0.9807) for $I=12^{+}$ and (0.8370,0.9722) for $I=14^{+}$.
As discussed in ref \cite{HertzZamick}. The $J=0^{+}$ $T=1$ matrix
element is not involved in these high spin states. The more relevant
comparison here is between the $J_{max}=9$ matrix element and the
one with $J=2$. With Q.Q the J=2 and J=7 matrix elements are 0.3482
and 0.4546. Hence J=2 is more attractive than J=9. In contrast with
CCGI {[}4{]} the respective numbers are 1.6500 and 0.9140. Hence with
CCGI the J=9 matrix element is more attractive than J=2 .

\section{Comparison with E($J_{max}$)}

\indent\indent Previously we had studied the overlaps of U$9j$
with wave functions of the E(J$_{max}$) interaction\cite{HertzZamick},i.e.
an interaction in which all two-body matrix elements vanish except
for the ones with $J=2j$. This interaction can only occur only between
a neutron and a proton. Studying this interaction gave us the idea
that the above set of U$9j$ coefficients could, in many cases, be
excellent approximations to wave functions that result from more realistic
interactions. Indeed the single U$9j$ $\langle(jj)^{9}(jj)^{J_{B}}|(jj)^{J_{p}}(jj)^{J_{n}}\rangle^{I}$,
with both $J_{p}$ and $J_{n}$ even is an exact eigenfunction of
E(9) for two protons and two neutrons in the $g_{9/2}$ shell. Indeed,
we get perfect overlap in the following cases with the $J_{max}$
interaction: $I=0$ $T=0$ $(J_{B}=9)$, $I=1$ $T=1$ $(J_{B}=8)$,
$I=2$ $T=1$ $(J_{B}=8)$ and $I=3$ $T=0$ $(J_{B}=7)$. Although
for $I=2$ $T=0$ a single U$9j$ is not an eigenstate, to an excellent
approximation the lowest two $I=2^{+}$ states have respectively $J_{B}=9$
and $J_{B}=7$. \\
 However, the E($J_{max}$) interaction taken by itself does not give
a reasonable spectrum. One of the sturdiest results in nuclear structure
is that all even-even nuclei have $I=0$ ground states. However with
an attractive E(9) interaction the lowest state has $I=16^{+}$. \\
 The Q.Q interaction has a much more reasonable spectrum, with an
$I=0^{+}$ ground state for even-even nuclei. A priori it is not clear
that single U$9j$'s could be reasonable approximations to the eigenfunctions
of Q.Q. The two particle matrix elements of this interaction are strongly
attractive not only for $J=J_{max}$ but also for $J=0$ (even more
so) and $J=1$. For this reason, it is more significant that the simple
sets of U$9j$'s noted above have high overlaps with wave functions
arising from Q.Q, at least for low spins. On the other hand, without
the study of the E($J_{max}$) interaction we would never have guessed
that a set of U$9j$'s existed which were close to wave functions
of realistic interactions.

\section{Closing Remarks}

In closing, we note that we have obtained an interesting result for
a system of two protons and two neutrons in a single $j$ shell. We
find that with the $J_{max}$ and Q.Q interactions states of low total
angular momentum are spectacularly well described by a single unitary
$9j$ coefficient. With more realistic interactions the overlaps,
though not quite as good are still greater than 0.9 in many cases.
It is of course well known that $9j$ coefficients are building blocks
from which could be used to construct wave functions but the surprise
here is that the unitary $9j$ coefficients are themselves in many
cases very close to the exact wave functions.\\
 Matthew Harper thanks the Rutgers Aresty Research Center for Undergraduates
for support during the 2014-2015 fall-spring session.

\subsection*{Table 1: Overlaps}

\centering $\begin{array}{|cc|c|c|c|cccccccccccccccccccccccccccccccccccccccccccccccccccccccccccccccccccccccccccccccccccccccccccccccccccccccccccccccccccccccccccccc|}
\hline\hline\hline\hline\hline\hline\hline\hline\hline\hline\hline\hline\hline\hline\hline\hline\hline\hline\hline\hline\hline\hline\hline\hline\hline\hline\hline\hline  &  & [\text{Q.Q U}9j] & \text{non-yrast} & [\text{CCGI,U}9j] & \text{non-yrast}\tabularnewline J & T & \text{Q.Qa} & \text{Q.Qb} & \text{CCGIa} & \text{CCGIb}\tabularnewline0 & 0 & 0.9996 &  & 0.9451 & \tabularnewline2 & 0 & 0.9999 &  & 0.9817 & \tabularnewline4 & 0 & 0.9986 &  & 0.9178 & \tabularnewline6 & 0 & 0.9871 &  & 0.8034 & \tabularnewline8 & 0 & 0.0481 & 0.9505 & 0.2315 & 0.0488\tabularnewline10 & 0 & 0.3635 & 0.8540 & 0.6399 & 0.7591\tabularnewline12 & 0 & 0.6590 & 0.5884 & 0.9806 & \tabularnewline14 & 0 & 0.8374 &  & 0.9722 & \tabularnewline16 & 0 & 1 &  & 1 & \tabularnewline2 & 1 & 0.9975 &  & 0.8870 & \tabularnewline4 & 1 & 0.9870 &  & 0.8850 & \tabularnewline6 & 1 & 0.9061 &  & 0.6319 & \tabularnewline8 & 1 & 0.0568 &  & 0.2027 & \tabularnewline10 & 1 & 0.3366 &  & 0.2711 & \tabularnewline12 & 1 & 0.7746 &  & 0.3916 & \tabularnewline3 & 0 & 0.9939 &  & 0.9912 & \tabularnewline5 & 0 & 0.9125 &  & 0.3329 & \tabularnewline7 & 0 & 0.7536 &  & 0.7607 & \tabularnewline9 & 0 & 0.3021 &  & 0.3925 & \tabularnewline1 & 1 & 0.9990 &  & 0.8369 & 0.5357\tabularnewline3 & 1 & 0.9974 &  & 0.9541 & \tabularnewline5 & 1 & 0.9634 &  & 0.9315 & \tabularnewline7 & 1 & 0.3120 &  & 0.6317 & \tabularnewline9 & 1 & 0.1055 &  & 0.1318 & \tabularnewline\end{array}$

\raggedright

\end{document}